\documentclass[aps,prx,amsmath, amssymb, english, ' twocolumn,reprint,floatfix, superscriptaddress]{revtex4-1}

\usepackage{multirow}
\usepackage{float}
\usepackage[normalem]{ulem}
\usepackage{textgreek}

\usepackage{bbm}
\usepackage{comment}
\usepackage{soul}

\usepackage{tikz}
\usepackage{microtype}
\usepackage{tkz-euclide}
\usepackage{tkz-graph}
\usepackage{epstopdf}
\usepackage{epsfig} 
\usepackage[applemac]{inputenx} 
\usepackage{amsmath}

\usepackage[unicode]{hyperref}
\hypersetup{
 colorlinks,
 citecolor=blue,
 filecolor=blue,
 linkcolor=blue,
 urlcolor=blue
}

\newcommand{\ket}[1]{\left|#1\right>}
\newcommand{\bra}[1]{\left<#1\right|}

\graphicspath{{./Images/}}

\date{}
\begin{document}

\title{Field-control of symmetry-broken and quantum disordered phases in frustrated moir\'e bilayers with population imbalance}
\author{Lorenzo Del Re} 
\affiliation{Max-Planck-Institute for Solid State Research, 70569 Stuttgart, Germany}
\author{Laura Classen} 
\affiliation{Max-Planck-Institute for Solid State Research, 70569 Stuttgart, Germany}
\affiliation{Department of Physics, Technical University of Munich, 85749 Garching, Germany}
\date{\today} 


\begin{abstract}
We determine the ground states and excitation spectra of the paradigmatic four-flavour Heisenberg model with nearest- and next-nearest-neighbor exchange couplings on the triangular lattice in a field controlling the population imbalance of flavor pairs. 
Such a system arises in the strongly correlated limit of moir\'e bilayers of transition metal dichalcogenides in an electric displacement field or in-plane magnetic field, and can be simulated via ultracold alkaline-earth atoms. 
We argue that the field tunes between effective SU(4) and SU(2) symmetries in the balanced and fully polarised limits and employ a combination of mean-field calculations, flavour-wave theory, and exact diagonalisation to analyse the intermediate, imbalanced regime. We find different symmetry-broken phases with simultaneous spin and excitonic order depending on the field and next-nearest-neighbor coupling.  
Furthermore, we demonstrate that there is a strongly fluctuating regime without long-range order that connects candidate spin liquids of the SU(2) and SU(4) limit. The strong fluctuations are facilitated by an extensive classical degeneracy of the model, and we argue that they are also responsible for a strong polarisability at 1/3 polarisation that survives from the mean-field level to the exact spectrum. 
\end{abstract}
\maketitle
\section{Introduction}
%
Strong correlations and frustration in quantum systems constitute a promising combination in the quest for useful phases of matter. 
They have a high potential for the realisation of unconventional spin orders with functional magnetic properties \cite{frustmag_2011}, as well as sought-after spin-liquid states with high entanglement and fractionalized excitations \cite{Balents2010,Savary_2017,doi:10.1146/annurev-conmatphys-031218-013401}. 
These phases are facilitated through a large degeneracy of the (classical) ground state which is why quantum spin models with various sources of degeneracy are intensely studied. 
For example, these include frustrated lattice geometries\cite{KITAEV20062,ReadSachd1991,sachdev1991large,Castelnovo2008,doi:10.1126/science.1064761,PhysRevB.106.125144,kuhlenkamp2022tunable}, higher SU(N) symmetries \cite{PhysRevB.65.214411,PhysRevB.77.134449,PhysRevB.84.054407,PhysRevB.84.174441,PhysRevLett.103.135301,PhysRevLett.121.097201,PhysRevB.98.195113,PhysRevResearch.2.013370,PhysRevB.100.024421,10.21468/SciPostPhys.8.5.076,PhysRevLett.125.117202,Savary2021}, 
or competing nearest- and next-to-nearest-neighbour interactions \cite{chubukov1992,joli1990,PhysRevB.91.014426,Zhu2015,Hu2015,PhysRevB.96.165141,Iqbal2016,doi:10.7566/JPSJ.83.093707,PhysRevB.96.075116,PhysRevB.94.121111,Ferrari2019,drescher2022dynamical,PhysRevLett.120.207203,wietek2023quantum}. 
For their controlled design and manipulation, tunable \cite{Scammell2023,scammell2023displacementfieldtunable} platforms are generally desirable, and even more so due to the fragile nature of quantum spin liquids. 
\begin{figure*}[htbp!]
    \centering
    \includegraphics[width = 1.99\columnwidth]{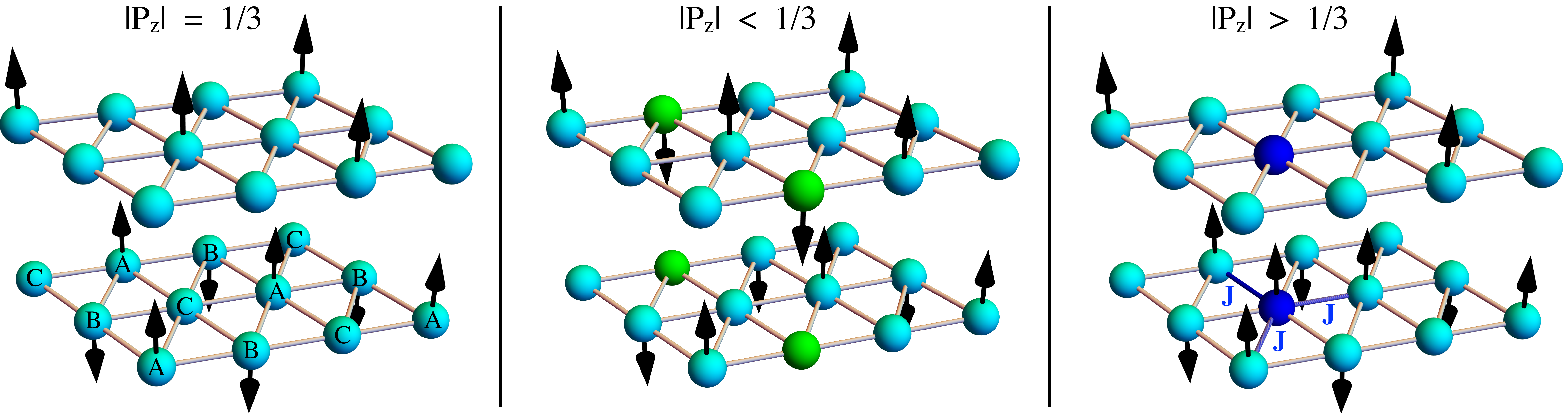}
    \caption{Sketch of different flavour configurations for $J^\prime = 0$ and    three values of the layer polarisation. We consider the realisation for the four flavours in moir\'e TMDs as an example and identify $\ket{1} = \ket{\uparrow \rm{t}}$, $\ket{2} = \ket{\downarrow \rm{t}}$, $\ket{3} = \ket{\uparrow \rm{b}}$, $\ket{4} = \ket{\downarrow \rm{b}}$ for $\uparrow$/$\downarrow$ spin and t(op)/b(ottom) layer. At zero field, the energy is minimised by any configuration with unequal nearest neighbours. (Left)  For an infinitesimal field, such degenerate, optimal configurations can be maintained while maximising the energy gain from the field when $|P_z| = 1/3$. A possible optimal configuration is given by a tripartite order where the A,B,C sublattices are completely polarised respectively with states $\ket{\uparrow \rm{b}}$, $\ket{\downarrow \rm{b}}$  and $\ket{\uparrow \rm{t}}$, and where the state 
    $\ket{\downarrow \rm{t}}$ is excluded. Flipping $\ket{\uparrow \rm{t}}\to\ket{\downarrow \rm{t}}$ randomly at 
    any site leaves the energy unchanged. (Center) Possible optimal configuration for $|P_z|<1/3$ obtained from the previous tripartite order where in two sites (highlighted in green) 
    states $\ket{\uparrow \rm{b}}$ or $\ket{\downarrow \rm{b}}$ have been flipped into $\ket{\downarrow \rm{t}}$. (Right) High-energy state ($3J$) with non-homogeneous polarisation obtained from the tripartite order by flipping at one site (highlighted in blue)  the state $\ket{\uparrow \rm{t}}$ into $\ket{\uparrow \rm{b}}$. }
    \label{fig:triangular_sketch}
\end{figure*}

The quantum simulation of strongly correlated fermions is established in ultracold atoms, and it was demonstrated that alkaline-earth atoms in optical lattices realise strongly correlated systems with a tunable  number of flavors N  and SU(N)-symmetric interactions  \cite{Gorshkov2010,PhysRevLett.105.190401,pagano2014,Cazalilla_2014,delre2018,tusi2022,Ibarra2021,taie2022observation}. Moir\'e transition metal dichalcogenides (TMD) offer recent solid-state alternatives for the controlled study of strongly correlated electron systems including triangular-lattice Hubbard models \cite{PhysRevLett.121.026402,PhysRevB.102.201115,PhysRevResearch.2.033087,PhysRevLett.121.266401,Regan2020,Tang2020,Jin2021,Shimazaki2020,Wang2020,Xu2020,Xu-Ping-Yao2021,Huang2021,kennes2021,ZangPRX2022,tscheppe2023magnetism,Motruk2023,Ros2023}. In particular, it is possible to form an SU(4) pseudo-spin out of layer and real spin degrees of freedom in twisted AB-stacked bilayers or three-layer hetero-structures with insulating middle layer \cite{Zhang2021}. In an experimental realisation with WSe${}_2$ competing electronic states with correlated insulators at integer fillings were reported \cite{Xu2022}. An important tuning parameter in these experiments is given by a perpendicular electric field which controls the layer polarisation. 
For integer layer populations, Mott insulators 
are formed at strong coupling, while at imbalanced layer population inter-layer excitonic insulators (EI) can emerge \cite{Xu2022,Zhang2022}. 
The Zeeman effect of an in-plane magnetic field and a population imbalance in cold-atom experiments acts 
analogously to such a polarising field. \
All these fields detune the population of pairs of flavours against each other. Hence,  they can be used to interpolate between effective SU(4) and SU(2) symmetric models 
 from balanced to full polarisation. This is particularly interesting for filling factors $n=1$ or $n=3$, where the SU(2) limit corresponds to the half-filled Hubbard model (as opposed to a band insulator for $n=2$). 
Theoretically, however, the effect of layer/population imbalance is not well studied. 

In this work, we investigate population-imbalanced AB-stacked TMD bilayers and ultracold fermionic alkaline-earth atoms via the SU(4) symmetric triangular-lattice Heisenberg model in a field. We map out the phase diagram as a function of the imbalance $P_z$ and next-to-nearest-neighbour coupling $J'$ employing flavour-wave theory and exact diagonalisation. In previous studies of the SU(2) symmetric case, a quantum spin liquid (of debated nature) was found between a 120$^\circ$ and a stripe magnetic phase when $J'$ is increased \cite{Iqbal2016,Ferrari2019,Zhu2015,Hu2015,drescher2022dynamical}. In the SU(4) limit, there is evidence for a transition from a quantum liquid to four-sublattice magnetic order upon increasing $J'$ \cite{Penc2003,Schrade2019,PhysRevResearch.2.013370,PhysRevLett.125.117202,Zhang2021}. We show that one can tune between these two limits via an external field and determine the different ground states and their excitations in between. 
We find that the SU(2) 120${}^\circ$ antiferromagnet (AFM) develops ferromagnetic ``dopants" in the minority layer and simultaneous 
tripartite inter-layer excitonic order when the field depopulates the half-filled majority layer. For larger $J'$, we obtain an evolution from the magnetic stripe order of the SU(2) limit 
into a four-sublattice state of the SU(4) limit with intermediate AFM and excitonic stripes.   
Furthermore, we demonstrate that a large part of the phase diagram is occupied by a strongly fluctuating phase (SFP) in which quantum fluctuations prevent any long-range order, and we argue that the SFP continuously connects the candidate spin liquids of the SU(2) and SU(4) limits.  
%
%
\section{The model}
We depart from the triangular-lattice Hubbard model with four flavours per site $\alpha = \{\ket{1},\ket{2},\ket{3},\ket{4}\}$. For concreteness, we identify these flavours with spin $\uparrow,\downarrow$ and layer (top, bottom) degrees of freedom in moir\'e TMDs \cite{Zhang2021}, i.e. $\ket{1} = \ket{\uparrow \rm{t}}$, $\ket{2} = \ket{\downarrow \rm{t}}$, $\ket{3} = \ket{\uparrow \rm{b}}$, $\ket{4} = \ket{\downarrow \rm{b}}$.
In the strong coupling limit and at fillings $n=1,3$ electron(s) per site, the Hubbard Hamiltonian is well captured by the Heisenberg model \cite{Kugel_1982}
\begin{align}\label{eq:Heis}
H = \sum_{ij} J_{ij}S^{\alpha}_\beta(i)S^{\beta}_\alpha(j) + \delta \sum_i \hat{P}_i^z,
\end{align}
where $S^{\alpha}_\beta = \ket{\alpha}\bra{\beta}$ and we consider nearest-neighbor (NN) and next-to-nearest-neighbor (NNN) super-exchange processes $J_{ij}$ of intensity $J$ and $J^\prime$, respectively.
In addition, we add the layer polarisation $\hat{P}^z = \ket{1}\bra{1} + \ket{2}\bra{2} - \ket{3}\bra{3} -\ket{4}\bra{4}$ 
so that a positive (negative) $\delta$ favors the population of the bottom (top) layer. Without loss of generality, we will consider the case of a positive $\delta$. 
By reshuffling flavours $\ket{1},\ldots,\ket{4}$, it becomes clear that the magnetic Zeeman term or a population imbalance of orbitals yield the same model Hamiltonian. 
If $\delta=0$, the Hamiltonian is SU(4) symmetric. 
For large $\delta$ the system approaches an effective SU(2) symmetry, where one layer is completely empty and the other half-filled. 

For later reference, we define observables of the system using the SU(4) generators via $\hat{O}^{ab} = \sum_{\alpha\beta} (\sigma_a\otimes \sigma_b)_{\alpha\beta} S_{\beta}^{\alpha}$, where $a = {0,1,2,3}$, with $\sigma_0 = \mathbbm{I}_{2\times 2}$ being the identity matrix and $\sigma_{(1,2,3)} = \sigma^{(x,y,z)}$  the Pauli matrices. Using this notation, the top (bottom) spin operators are given by $\hat{S}_{t (b)}^{k} = 1/2[\hat{O}^{0,k} + (-) \,\hat{O}^{3,k}]$, with $k =1,2,3$. Inter-layer processes, which define the excitonic order parameter, are encoded by 8 operators $\hat{O}^{a,b}$ with $a = {1,2}$ and $b = {0,1,2,3}$.
\begin{figure*}
     \centering
     \includegraphics[width = 1.99\columnwidth]{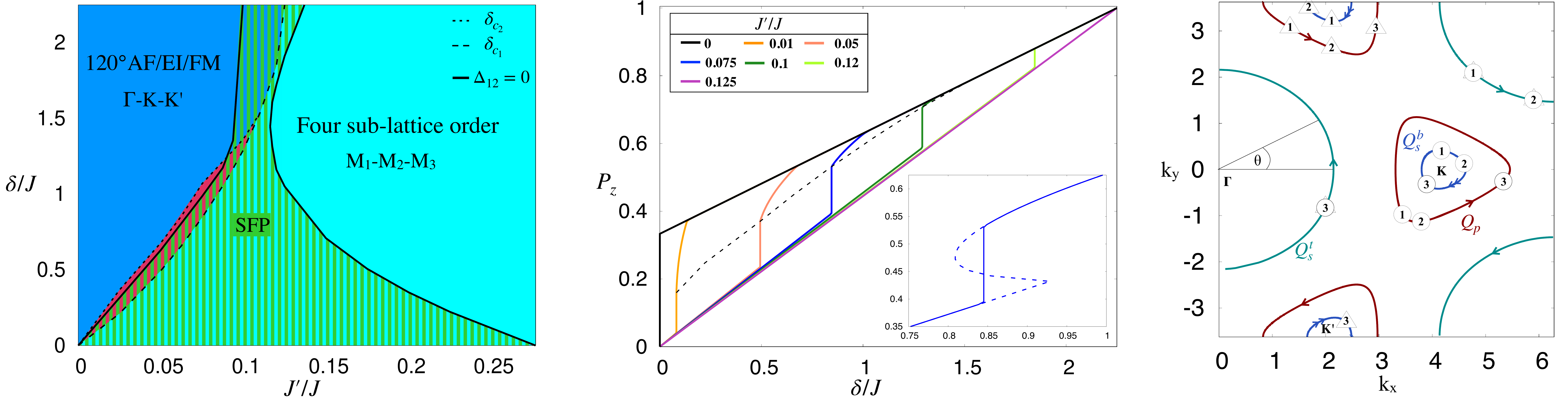}
     \caption{\small (Left)  Phase diagram in the plane $\delta$ vs $J^\prime$ within mean-field (dashed lines $\delta_{c_1}$ and $\delta_{c_2}$) and including quantum fluctuations (solid lines). We find three phases displaying long-range order which we label with the wave-vector triplet $Q_s^t$, $Q_p$, $Q_{s}^b + Q_p$ (see text). The blue $\Gamma$-K-K$^\prime$ regime describes $120^\circ$ spin (AF) and exciton order (EI) with minority ferromagnetism (FM). In the turquoise M$_1$-M$_2$-M$_3$ region, spin and excitons form stripes in each layer leading to a four-sublattice order, and in the red region, they order with incommensurate wave vectors. The green area in the phase diagram that we named strongly fluctuating phase (SFP) indicates the region where quantum fluctuations suppress the order parameter $\Delta_{12}$ to zero. It connects candidate spin liquids of the SU(2) ($\delta/J\gg1$) and SU(4) ($\delta=0$) limit. Vertical blue/turquoise/red hatches refer respectively to the $\Gamma$-K-$K^{\prime}$/ M$_1$-M$_2$-M$_3$/ incommensurate orders predicted by mean-field theory that are suppressed by quantum corrections.
     (Center) Mean-field polarisation curves as a function of the external field for different values of $J^\prime/J$. The inset shows the Maxwell construction for $J^\prime/J = 0.075$. Dashed line is a guide to the eye for the first-order jump. 
(Right) Ground state degenerate manifold  for $|P_z | = 0.183 < 1/3$ given by three different curves made of Q-vector triplets $Q_s^t$ (green), $Q_s^b$ (blue), and $Q_p$ (red) that minimize the classical energy.  
We explicitly mark three examples of triplets $(Q_s^t(n),Q_s^b(n),Q_p(n))$ numbered by $n=1,2,3$. States with $Q_s^b,Q_p$ around K (circles) and K$^\prime$ (triangles) are degenerate.  
Arrows denote how triplets evolve with $\theta$ accounting for double valued $Q_s^b(\theta + \pi;K^{(\prime)}) =Q_s^b(\theta;K^{(\prime)}) $.
}
     \label{fig:phase_diagram}
 \end{figure*}
\section{The Classical Ground State}
We first determine the mean-field phase diagram as function of $J'$ and $\delta$ and consider the role of quantum fluctuations 
in the next section.  
In the SU(4) limit $\delta=0$ and for $J'=0$ the classical ground state (GS) is extensively degenerate: any state with different flavours on neighbouring sites minimises the energy. A finite $J'$ selects a four-sublattice ground state out of this manifold \cite{Penc2003}. Similarly, at $J^\prime = 0$, we expect an infinitesimal field $\delta$ to select a three-sublattice state out of the manifold because it possesses the maximal polarisation $|P_z|=1/3$ (see Figure \ref{fig:triangular_sketch}).  
The three-sublattice state with $|P_z|=1/3$ is still extensively degenerate because any site of the third sublattice can be spin up or down.  When $|P_z|<1/3$ the ground state keeps being highly degenerate and can be obtained starting from the three-sublattice state by substituting e.g. flavors $3$ or $4$ with flavors $2$ or $1$ with the  constraints of always having different flavours on neighboring sites, as shown in Figure \ref{fig:triangular_sketch}. 
In the 
 SU(2) limit for large polarisation $\delta/J\gg 1$, the effective half-filled triangular lattice possesses $120^\circ$ antiferromagnetic order for small $J'$ and transitions to a stripe phase for $J'>1/8$ {\cite{joli1990,chubukov1992}}. 
\par To obtain the mean-field phase diagram for general $\delta$ between these limits, we perform a product state ansatz $\ket{\Psi} = \prod_i\ket{\psi_i}_i$ and minimize the classical energy $E_{cl} =\bra{\Psi}H\ket{\Psi} $ \cite{Joshi1999}. As an ansatz we choose a state with in-plane spin order and homogeneous layer polarization $P_z$ given by 
$\ket{\psi_i} =  \frac{\sqrt{1+P_z}}{2}\left(\ket{1} + e^{iQ_s^t\cdot R_i}\ket{2}\right) 
    +e^{i Q_p\cdot R_i}\frac{\sqrt{1-P_z}}{2}
    \left(\ket{3} + e^{i Q_s^b\cdot R_i }\ket{4}\right)$,
where $Q_p$, $Q_s^b$ and $Q_s^t$ are respectively the wave-vectors associated with the relative modulations of the pseudo-spin (layer), bottom spin and top spin. 
We choose this ansatz $\ket{\psi_i}$ because it can interpolate between the limiting cases and because it has lower energy than states with flavour-polarised sites (see below).  
With this ansatz we can show analytically that for almost complete polarisation $|P_z|\lesssim1$, the ground state to order $\mathcal O ((1-|P_z|)^2)$ is given by a configuration $Q_s^t = \Gamma$ and $Q_s^b = Q_p = K$ (see Appendix \ref{appendix:classical}). 
This describes ferromagnetic (FM) order of the top spin in the scarcely populated layer, and tripartite 120$^\circ$ order of the bottom spin in the densely populated  layer, which reduces to the conventional $120^\circ$ AFM order in the SU(2) limit. Simultaneously, we have a four-component inter-layer excitonic order parameter, where the non-vanishing components are given by  $\hat{O}^{1,0}$,  $\hat{O}^{2,3}$, $\hat{O}^{1,1}$, $\hat{O}^{1,2}$. The pairs  ($\hat{O}^{1,0}$,  $\hat{O}^{2,3}$) and  ($\hat{O}^{1,1}$, $\hat{O}^{1,2}$)  also form a 120$^\circ$  configuration. 
At $P_z=1/3$, the tripartite $\Gamma K K$ state with homogeneous polarisation is degenerate with the manifold of three-sublattice state with flavour-polarised sites. But for increasing $|P_z|$, the energy of the $\Gamma K K$ state remains minimal $\mathcal O ((|P_z|-1/3)^2J)$, while the energy of the flavour-polarised state rises strongly $\mathcal O (J)$ (see right panel of Figure \ref{fig:triangular_sketch}).

Our numerical minimisation confirms the analytical considerations and allows us to obtain the full mean-field phase diagram between limiting cases. 
 In total, we find three broken-symmetry states divided by two critical lines $\delta_{c_1}$ and $\delta_{c_2}$ (see Fig.~\ref{fig:phase_diagram}). 
For values of the field $\delta < \delta_{c_1}$ or $J'/J>1/8$ the energy is minimised by the wave-vector triplet $Q_s^t = Q_s^b = M_1$ and $Q_p = M_2$ (and those related to this by symmetry). 
This gives rise to a four-sublattice order where top and bottom spin $\hat S_{t,b}$ form the same stripe arrangement in one lattice direction given by $M_1$, and the inter-layer exciton  has two components
$\hat O^{1,0}$ and $\hat O^{1,1}$ forming stripes in the other lattice directions defined respectively by $M_2$ and $M_3$. 
This four-sublattice order is reduced to stripes in the SU(2) limit for $J'>1/8$ \cite{joli1990,chubukov1992} 
and recovers the SU(4)-symmetric case at $\delta = 0$ \cite{Penc2003}. 
When $\delta > \delta_{c_2}$ the GS is determined by the tripartite $\Gamma K K$ state. 
For $\delta_{c_1}<\delta <\delta_{c_2}$ we find a third GS with triplets of incommensurate wave vectors.
The incommensurate order is classically stable for $J^\prime/J \lesssim 0.11$. 

For further characterisation, we calculate the layer polarisation as function of the external field $\delta$ for different values of the NNN exchange intensity $J'$ (see Fig.~\ref{fig:phase_diagram}). 
In the four-sublattice state the polarisation increases linearly as a function of the field  $|P_z| = \delta/[2(J+J^\prime)]$ either until full polarisation if $J'/J>1/8$ or until  $\delta = \delta_{c_1}$, where it jumps abruptly to a higher value and the order becomes tripartite (if $0.11\lesssim J'/J<1/8$) or incommensurate (if $J'/J\lesssim 0.11$). 
In the incommensurate phase the polarisation has a nonlinear behavior for $\delta_{c_1}<\delta<\delta_{c_2}$ and at $\delta = \delta_{c_2}$ the system continuously transitions into the tripartite phase. 
In the tripartite phase the polarisation increases linearly as $|P_z| = (8/27)\delta/J + 1/3$.
We evaluate the size of the first-order jump using Maxwell's construction (see Figure \ref{fig:phase_diagram}), which displays a non-monotonic behaviour as function of $J'$. 
At $J^\prime = 0$ the polarisation jumps at $\delta = 0^+$ from zero to $|P_z| = 1/3$. In approaching this point, the slope of the polarisation in the incommensurate region becomes steeper with decreasing $J^\prime$ 
so that the layer polarisability $\kappa = \frac{d P_z}{d\delta}$ diverges at $J^\prime = 0$. Such an instability of the incommensurate order is accompanied by the onset of a continuous manifold of degenerate 
ground states for the classically forbidden values of the polarisation $|P_z| <1/3$ \footnote{We note that,  differently from the states sketched in Figure \ref{fig:triangular_sketch} for $|P_z|<1/3$, these degenerate states yield homogeneous polarisation.}.  
It is characterised by Q-vector triplets defined on three curves $Q_s^t(\theta)$, $Q_s^b(\theta)$, $Q_p(\theta)$ around $\Gamma$ and K (K$^\prime$) in the Brillouin zone parameterised by the angle $\theta$ as shown in Fig.~\ref{fig:phase_diagram} for the representative case of $|P_z| = 0.183$. 
%
\section{The role of quantum fluctuations}
We study the excitation spectrum and the stability of the different phases within flavour-wave theory. 
Before adding quantum fluctuations to the mean-field solution, it is useful to introduce a unitary transformation $\mathcal{U}_i$ that brings $\ket{\psi_i}\to\ket{1}$, in the same spirit as it has been done for evaluating the spin waves of the 120 degrees AFM \cite{Cherny2009} and the non-homogenous quantum Ising model \cite{delre2016}. In this new basis, the ground state assumes the form of a homogeneous fully polarized state where every site is in the state $\ket{1}$. 
The Hamiltonian in the new basis reads 
\begin{align}\label{eq:Heis_unit}
    \mathcal{U}\,H\,\mathcal{U}^\dag =&\sum_{ij}\sum_{\alpha \alpha^\prime \beta \beta^\prime}J(\tau)\, \kappa_{\alpha \alpha^\prime}(\tau)\kappa^*_{\beta \beta^\prime}(\tau) S^{\alpha^{\,}}_{\beta^{\,}}(i)\,S^{\beta^\prime}_{\alpha^\prime}(j) \nonumber \\ &+ \delta\sum_i\sum_{\alpha\beta} \tilde{P}^{z}_{\alpha\beta}S^{\alpha}_\beta(i), 
\end{align}
where $\kappa_{\alpha\beta}(\tau) = \bra{\alpha} \mathcal{U}_i\,\mathcal{U}^\dag_j\ket{\beta} $,  with $\tau = R_i-R_j$, and  $\tilde{P}^z_{\alpha\beta} = \bra{\alpha}\mathcal{U}_i \,\hat{P}^z_i\,\mathcal{U}^\dag_i\ket{\beta}$.
We rewrite the Hamiltonian in Eq.(\ref{eq:Heis_unit}) as $E_{cl} + \delta H$ where $\delta H$ contains the quantum fluctuations in terms of a generalised Holstein-Primakoff transformation $S^1_1=M-\sum_n b_n^\dagger b_n$, $S^1_m=b_m\sqrt{M-\sum b_n^\dagger b_n}$, and $S^n_m=b_n^\dagger b_m$ with three bosonic operators $b_n^{(\dagger)}$, $n\in\{1,2,3\}$, and expansion in $1/M$. 
In the harmonic approximation, we can calculate the characteristic flavour-wave spectrum for the different phases by diagonalising $\delta H$ for various values of $\delta$ and $J'$. 
We also explicitly checked that the dynamical structure factors contain the same excitations (see Appendix \ref{appendix:dynamical}). 

\begin{figure}
    \centering
    \includegraphics[width = \columnwidth]{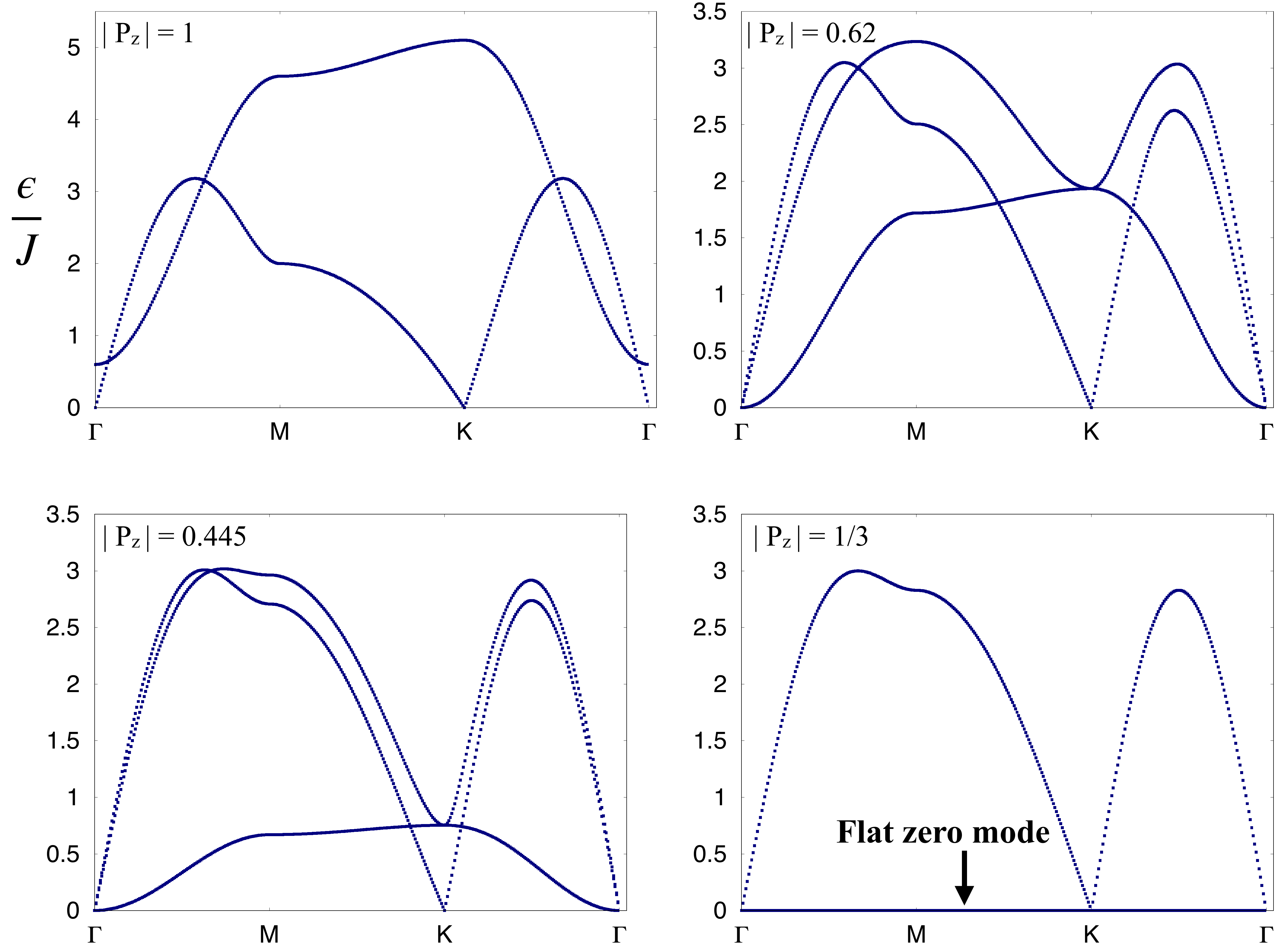}
    \caption{Evolution of the 
    flavour-wave spectrum plotted as a function of the crystalline momentum for different values of the 
layer polarization $|P_z| = 1, 0.62, 0.445,1/3.$ }
    \label{fig:qf}
\end{figure}
We  first discuss the results at $J^\prime = 0$.
In Figure \ref{fig:qf}, we show four different spectra  as a function of the crystalline momentum for different values of the polarization greater than 1/3. For large enough values of $\delta$, when the system is fully polarized $|P_z| = 1$, the spin waves of the half-filled bottom layer coincide with the ones of the mono-layer 120$^\circ$ AFM  \cite{joli1989,joli1990,chubukov1992,Cherny2009}. They are gap-less at  $k =\Gamma, K$   and their energy increases linearly in the vicinity of those points. On top of these excitations, we find two degenerate gapped ``ferromagnetic"-like spin waves, which encode the hopping of one electron from the bottom to the top layer. 
Their gap at the $\Gamma$ point is controlled by $2|\delta - \delta_s|$ for $\delta > \delta_s$, where $\delta_s = 9/4 J$ is the minimum value of the field necessary for full polarization.  For intermediate values of $|P_z|$ lying in the interval $(1/3,1)$, we have three distinct Goldstone modes at the $\Gamma$ point, two of which have a linear dispersion with different velocities while the third one  displays a quadratic behavior that is associated to the FM $Q_s^t = 0$ wave-vector of the top-layer spin order. At the $K$-point we observe only one linear Goldstone mode. Furthermore, the FM mode is strongly suppressed by decreasing the polarization toward the critical value of $1/3$. 
At the same time, the two other branches approach each other.  
At the critical point $|P_z| = 1/3$, we obtain two degenerate excitations with linear Goldstone modes at the $\Gamma$ and $K$ points, and the FM mode completely flattens to zero. 
As mentioned before, these zero-energy excitations are already present in the classical picture, where any spin in the minority layer of the state with flavor-polarised sites can be flipped without energy cost, and we observe that they survive upon inclusion of Gaussian quantum fluctuations.
They strongly affect the system's properties and mark the end of long-range order 
that for $J^\prime = 0$ is stable up to $|P_z| = 1/3$.

\begin{figure}
    \centering
    \includegraphics[width = \columnwidth]{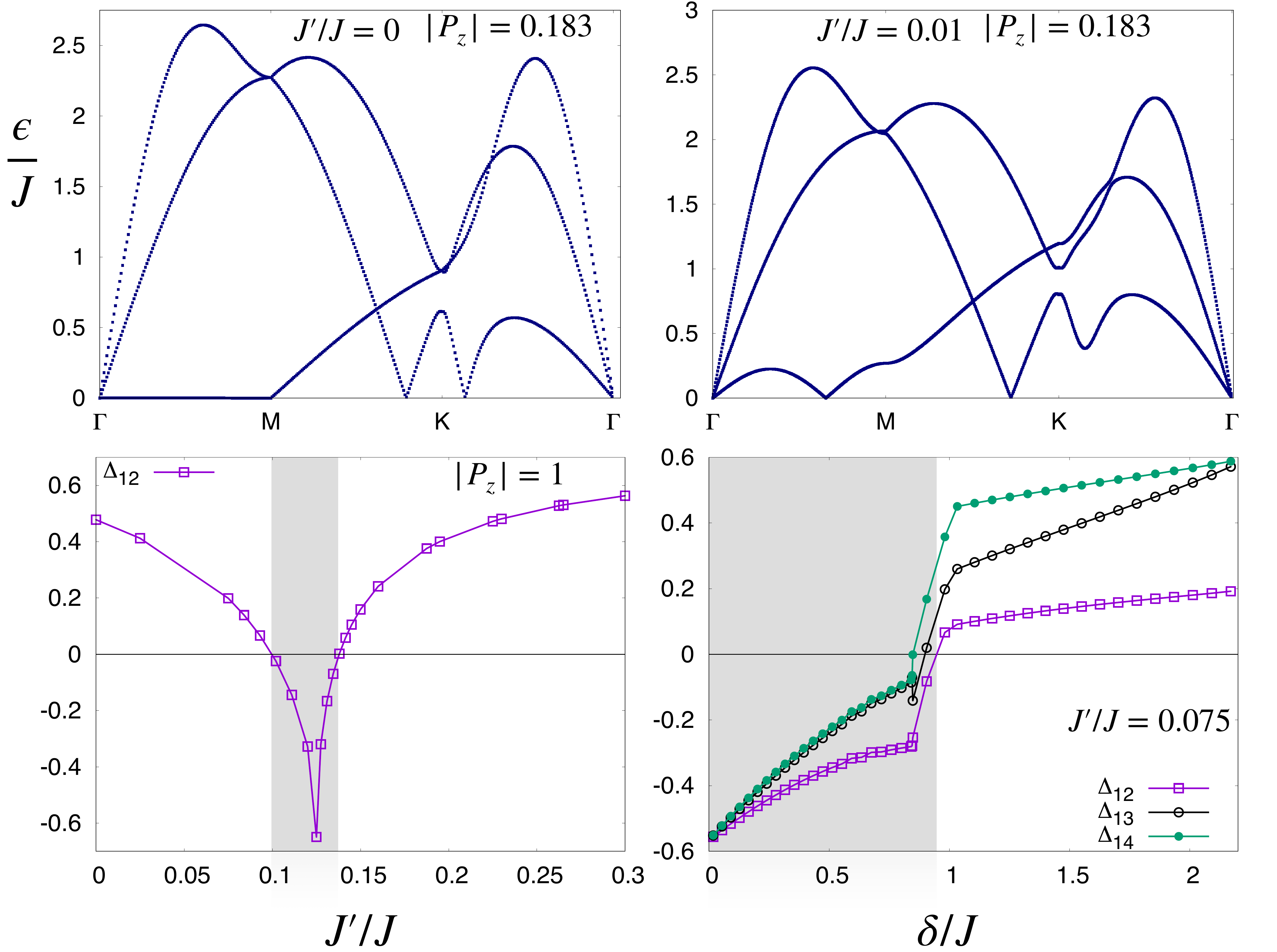}
    \caption{(Top Panels) Flavour-wave spectra for $|P_z| = 0.183$ and two different values of $J^\prime/J = 0, 0.01$. (Lower-left) Renormalised order parameter as a function of $J^\prime/J$ at full polarisation. (Lower-right) Order parameter components as a function of the external field for fixed $J^\prime/J = 0.075$. When $\Delta_{1\alpha}<0$ (gray shaded area in the plots), quantum fluctuations destroy the mean-field order.}
    \label{fig:qf_inc}
\end{figure}

It is also instructive to analyse the role of quantum fluctuations in the classically forbidden region of the parameter space $|P_z|<1/3$.
In this region, the classical ground state is highly degenerate. However, quantum fluctuations can remove such a high degeneracy and select particular states which become energetically favored once quantum corrections are taken into account, a well known phenomenon  known as quantum order-by-disorder \cite{villain1980order,Zhit2012,toth2010,Bauer2012,Jack2015}.
In order to determine the ground state, we calculate quantum corrections to the  energy for every degenerate classical state by evaluating the zero-point energy of quantum fluctuations (see Appendix \ref{appendix:order_by_disorder}). 
We find that $Q_s^t$ and $Q_s^b$ selected by quantum fluctuations lie respectively on the $\Gamma$-$M$ and $K$-$M$ directions. 
Figure \ref{fig:qf_inc} displays the flavour-wave spectrum as a function of the crystalline momentum for $P_z = -0.183$. We observe that the fluctuation energy vanishes along the $\Gamma$-$M$ direction and that we have Goldstone modes at the M-point and at two incommensurate wave vectors  lying respectively on the $M$-$K$ and $K$-$\Gamma$ directions. The nodal line in $\Gamma$-$M$ direction is again a consequence of the high degeneracy of the classical ground state and disorders the system. As we show below the presence and proximity of the zero modes strongly renormalises the order parameter and suppresses it to zero in large parts of the phase diagram. 
\\ \par 
When $J^\prime$ is finite the degeneracy of the classical ground state is removed everywhere in the phase diagram even in the case of incommensurate order. Figure \ref{fig:qf_inc} shows the flavour-wave spectra for $J^\prime/J = 0.01$ and $|P_z|= 0.183 $. We find that even small values of the NNN super-exchange are enough to remove the nodal lines appearing in the classically forbidden region at $J^\prime = 0$. 
Furthermore, we observe that the Goldstone mode appearing at the $M$-point for $J^\prime = 0$ now shifts to an incommensurate vector lying in the $\Gamma$-$M$ direction, and that the extra Goldstone mode appearing 
along $\Gamma-K$ for $J^\prime = 0$  acquires a gap. 

The regularisation of the spectra introduced by a non-vanishing $J^\prime$ allows us to quantify the impact of quantum fluctuations on the order parameter and map out the phase diagram beyond the mean-field approximation. 
The classical ground state in the new basis defined in Eq.(\ref{eq:Heis_unit}) is given by the completely polarised state $\prod_i\ket{1}_i$, with $n_1 = 1$ and $n_2 = n_3 = n_4 = 0$, where $n_\alpha = \frac{1}{V}\sum_i\left<S^\alpha_\alpha(i)\right>$ is the $\alpha$-th flavor population. With the inclusion of quantum fluctuations, the density matrix $n_{\alpha\beta}= \frac{1}{V}\sum_i\left<S^\alpha_\beta(i)\right>$ acquires off-diagonal terms and it has a block-diagonal form given by a 1$\times$1 ($\alpha,\beta=1$) and a 3$\times$3 ($\alpha,\beta\in\{2,3,4\}$) block (see Appendix  \ref{appendix:inclusion}). 
Then, it is natural to choose the basis that diagonalises the the density matrix. The eigenvalues are the occupation numbers of the minority flavors that, for simplicity, we still refer to as $n_2$, $n_3$ and $n_4$, and we sort in descending order, i.e. $n_2> n_3> n_4$. 
The population of the majority flavor can be computed from the knowledge of the renormalised minority ones via $n_1 = 1 - \sum_{\alpha = 2}^4 n_{\alpha}$. 
In the  generic case,  the four occupation numbers are non-vanishing and we can express the order parameter in terms of three components $\Delta_{1\alpha} = n_1 - n_\alpha$ with $\alpha = 2,3,4$. We identify the region in the phase diagram enclosed by the contour   $\Delta_{12} = 0$ (see Figure \ref{fig:phase_diagram}) as a strongly fluctuating phase (SFP), where quantum fluctuations are so strong that  $n_1$  does not represent the occupation of the majority flavours anymore \footnote{This is not the only possible choice to define the strong fluctuations region in the phase diagram. Sometimes $n_1 = 0$ is chosen to mark the onset of possible spin-liquid phases. However, we found that our choice is applicable to the SU(2)-limit without any ambiguity. In fact, for spin $1/2$, $n_1 = 0$ would yield a negative expectation value of the order parameter instead of zero, and therefore such a choice would underestimate the role of quantum fluctuations. }.

The lower panels of Figure \ref{fig:qf_inc}  show the order parameter $\Delta_{1\alpha}$  as a function of NNN super-exchange and polarisation. As we explained before,  at $|P_z| = 1$ we have a fully polarised layer with one fermion per site, which is equivalent to the SU(2) Heisenberg model. In this case, the GS transitions from a 120$^\circ$ AFM to a striped phase at $J^\prime/J = 1/8$ on the mean-field level \cite{joli1990,chubukov1992}. We note that the occupation numbers $n_3 = n_4 = 0$ in the SU(2) limit, and the order parameter has only one component,  $\Delta_{12}$. 
It displays a cusp at the transition point $J^\prime/J = 1/8$ and crosses zero at two different close-by points, namely $J_{a}^\prime/J \approx 0.1$ and  $J_b^\prime/J \approx 0.138$.  
Between these points $J_a<J^\prime<J_b$ the order parameter vanishes and becomes negative. This means that the harmonic approximation cannot be trusted anymore because quantum fluctuations are so strong to the point of destroying the order parameter. Interestingly, this interval is quantitatively comparable to the range of values where a spin-liquid phase was predicted by Monte-Carlo \cite{Iqbal2016,Ferrari2019} and DMRG  \cite{Zhu2015,Hu2015} calculations. 
\begin{figure*}[htbp!]
    \centering
    \includegraphics[width = 1.95\columnwidth]{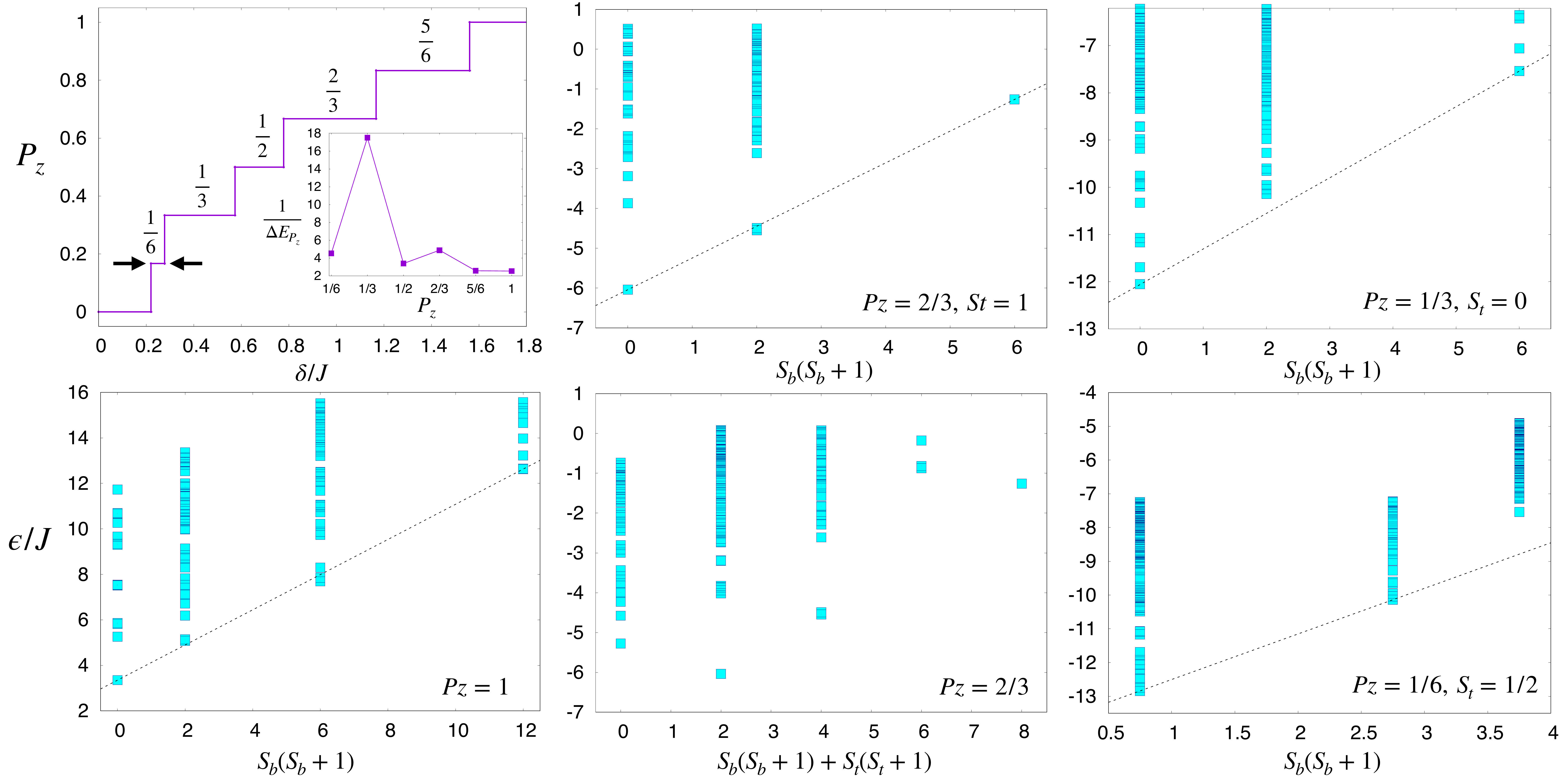}
    \caption{Numerical results from Lanczos Exact Diagonalisation.  (Left top) Polarisation as a function of the external field. The inset displays the inverse of the plateaux width as a function of polarisation, which is an estimate of the system's polarisability (see main text). The other panels show the low-energy eigenspectra of the Hamiltonian belonging to different polarisation sectors. Antiferromagnets form the  Anderson tower of states proportional to $S^2/N$. Dashed lines are guide to the eyes.}
    \label{fig:exact-spectra}
\end{figure*}

As an example, we also show the occupation numbers as a function of the external field at fixed $J^\prime/J = 0.075$ in Figure \ref{fig:qf_inc}. For values of $P_z$ close to complete polarisation, the system is in the $\Gamma$-K-K phase and we 
find $n_4 = 0$ 
so that $n_1=\Delta_{14}$  only mildly decreases as a function of 
 decreasing $\delta$. At $\delta \sim J$, the GS becomes incommensurate, $n_4> 0$ and 
 all $\Delta_{1\alpha}$ are lowered quite rapidly  by decreasing $\delta$ crossing 
 zero ($\Delta_{1\alpha} = 0$) thereafter within the incommensurate phase. 
For smaller values of $\delta$, the harmonic approximation breaks down and the system enters the strongly fluctuating regime. 

We summarize the impact of quantum fluctuations in the phase diagram in Fig.~\ref{fig:phase_diagram}. 
We observe that the strongly fluctuating regime connects the putative spin liquid phases of the SU(4) and SU(2) limits when the layer population is varied with $\delta$. Similarly, for larger $J'$ the four-sublattice and stripe phases of the limiting cases are continuously connected. In contrast,  the $\Gamma KK$ phase, which reduces to the $120^\circ$ AFM for full polarisation, is destroyed by strong fluctuations before the SU(4) $\delta=0$ limit can be reached. For small $J'/J$ the strongly fluctuating regime is preempted by an incommensurate phase. 
\section{Exact Spectra}
In order to gain  additional insights about the underlying physics of the exact solution we perform Lanczos exact diagonalisation for a 12-site cluster with periodic boundary conditions (PBC) defined by the lattice vectors $T_1 = 4a_2 - 2 a_1$ and $T_2 =2 a_1 + 2 a_2 $, where $a_1 = (1,0)$ and $a_2 = \frac{1}{2}(1,\sqrt{3})$.  In Fig.~\ref{fig:exact-spectra} we show  the polarisation as a function of the external field.  
The finite size 
allows for seven different polarisation values, namely $P_z = 1,5/6,2/3,1/2,1/3,1/6,0$. Therefore, the polarisation 
 displays seven plateaus. 
The width of a given plateau matches the energy difference between the two ground states belonging to the two different polarisation sectors, i.e. $\Delta E_{P_z} = E_{0,P_z} - E_{0,P_z-1/6}$. In the thermodynamic limit, the layer-polarisability is given by $\kappa 
=(\lim_{\Delta{P_z} \to 0}\frac{\Delta E_{P_z}}{\Delta{ P_z}})^{-1}$. Hence, we can use the finite difference $\Delta E_{P_z}$ as an approximate estimate of the  inverse layer polarisability (see inset in Fig.~\ref{fig:exact-spectra}).  We observe that $\Delta E_{P_z}^{-1}$ is strongly peaked at $P_z =1/3$, where its value 
is  one order of magnitude larger than for the other polarisations. 
This is in line with the sharpening of the slope of the polarisation upon approaching $|P_z|=1/3$ in the mean-field calculation (Fig.~\ref{fig:phase_diagram}). 

Further insights about the tendency of the system to stabilise a long range order can be gained by studying the low-energy sector of the energy eigenspectrum. Mean-field theory plus quantum fluctuations suggest that the bottom (top) layer orders antiferromagnetically (ferromagnetically) for $P_z \geq 1/3$. We can check the tendency of the system to form an AFM order in the bottom layer by plotting the energy eigenvalues as a function of the total spin of the bottom layer, i.e. $S_b^2 = (S_{b,\text{tot}}^x)^2 + (S_{b,\text{tot}}^y)^2 + (S_{b,\text{tot}}^z)^2$, with $S^{(x,y,z)}_{b,\text{tot}} = \sum_i S^{(x,y,z)}_{b}(i)$, which is a conserved quantity. In fact,  the low energy eigenstates  of antiferromagnets form a structure  known as the Anderson tower of states: well separated states that are proportional to $S^2/N$ \cite{Anderson1952t,Bernu1994,Penc2003,Aron2019}. On the other hand, ferromagnets do not display the same behavior. 
Therefore, for $P_z\geq 1/3$, we expect  the  spectrum to only form the Anderson tower when the energy eigenvalues are plotted as function of  $S_b^2$, and not when it is plotted as a function of the total spin.  In Figure \ref{fig:exact-spectra}, we show the energy spectra for different values of the polarisation sectors. For $P_z=1$, the spectrum of the bottom layer reproduces the one of the SU(2)-Heisenberg model \cite{Bernu1994}. At $P_z=2/3$, the ground state is found in the sector with $S_b = 0$ and $S_t = 1$: we observe that the low-energy excited states with $S_t = 1$ show a linear behavior as a function of $S_b(S_b +1)$.  However, when the spectrum is plotted as a function of the total spin eigenvalues the linear behavior is lost as expected. Furthermore, the GS is found in the maximal spin sector of the top layer, signalling the tendency towards a FM order in the top-layer.
At $P_z = 1/3$, we start to observe a smalldeviation from linear behavior of the spectrum as a function of $S_b(S_b+1)$. 
Furthermore, we now find the GS 
in the minimal top-layer spin sector, i.e. $S_t =0$, probably suggesting that FM order is already lost for this value of the polarisation.
At $P_z = 1/6$ we observe a sizable deviation from linear behavior, suggesting that at this value of the polarisation the bottom layer does not order antiferromagnetically.   
Thus, the results of the exact diagonalisation provide strong indication that the ground state is quantum disordered for small enough polarisations. 
%
\section{Conclusions}
Motivated by realisations in cold atoms and moir\'e TMDs, we carried out a theoretical study of the paradigmatic 
SU(4)-Heisenberg model on the triangular lattice in presence of a polarising field $\delta$, which controls a population imbalance of of flavor pairs. On the classical level, the model is strongly frustrated with an extensive ground-state degeneracy, which we argued can persist for finite fields up to polarisations $|P_z|\leq 1/3$.  
Through a combination of variational mean-field calculations, flavour-wave theory, and exact diagonalisation we determined the  ground states 
and excitation spectra for different values of the field $\delta$ and nearest-neighbor coupling $J^\prime$. 
We mapped out a rich phase diagram with commensurate and incommensurate long-range orders, as well as a strongly fluctuating phase that shows evidence for a quantum disordered ground state. 

For small $J^\prime/J$ and large enough $\delta$, we found a tripartite phase where the bottom-layer spin and the inter-layer exciton order in a 120$^\circ$ fashion while the top layer is ferromagnetically ordered. 
Accordingly, its flavour-wave spectrum shows AFM and FM excitations. 
For large $J'/J$, we found a four-sublattice phase for all values of $\delta$ which displays a striped configuration for top spin, bottom spin, and excitonic order parameter. 
In between these two phases, we found a small sliver of incommensurate order and a large regime of a strongly fluctuating phase where quantum fluctuations suppress long-range order. This is mirrored by extended zero modes in the flavour-wave spectrum for $J'\rightarrow 0$. 
Our ED calculations on a 12-site cluster provided supporting evidence for a transition from an ordered to a disordered phase between large and small values of the polarisation. Furthermore, we observed a strong increase in the polarisability upon approaching the transition. 

We argued that in the case of full polarisation, where the system effectively models the SU(2)-symmetric triangular lattice, the SFP can be identified as the precursor of the spin liquid state found in DMRG and Monte Carlo calculations \cite{Iqbal2016,Ferrari2019,Zhu2015,Hu2015,drescher2022dynamical}. Similarly, the SFP coincides with a putative quantum liquid in the SU(4) limit $\delta=0$ \cite{PhysRevResearch.2.013370,PhysRevLett.125.117202,Zhang2021}. Interestingly, the two limits are  continuously connected via the SFP making future studies on possible liquid phases in this regime highly desirable. This could also shed new light on the debated nature of the SU(2) quantum spin liquid. 

Given the 
 rich phase structure, spin and charge configurations
can be effectively 
manipulated via the external field. 
The tunability of cold-atom and moir\'e systems provides an ideal opportunity to investigate these quantum many-body phases experimentally.  The polarising field can be readily controlled and the NNN exchange $J'$ is not expected to be very large so that it seems possible to reach the SFP regime, in particular, because it occupies an extended region of the phase space. For example, the softening of the FM mode, that could be probed by measuring the dynamic structure factor (see Appendix \ref{appendix:dynamical}), would yield direct evidence of approaching the SFP.  
Intriguingly, a recent experiment on twisted AB-stacked WSe${}_2$ reported evidence for paramagnetic insulators at hole density $n=1$ for zero and full polarisation, and a potential excitonic insulator for intermediate polarisation. While this 
is in accordance with our findings, further experimental and theoretical investigations are needed to elucidate the nature of the insulating states. 
Most notably, this includes the possible emergence of Mott insulators in the associated Hubbard model. 
\section{Acknowledgments}
We thank Massimo Capone, Adriano Amaricci, Elio K\"onig, Michael Knap, Johannes Knolle, Frank Pollmann, and Thomas Sch\"afer for valuable discussions. 

\appendix
\section{Classical energy in limiting cases}\label{appendix:classical}
The explicit expression for the classical energy $E_{cl} =\bra{\Psi}H\ket{\Psi} $ using the product state ansatz $\ket{\Psi} = \prod_i\ket{\psi_i}_i$ with $\ket{\psi_i} =  \frac{\sqrt{1+ P_z}}{2}\left(\ket{1} + e^{iQ_s^t\cdot R_i}\ket{2}\right) 
    +e^{i Q_p\cdot R_i}\frac{\sqrt{1- P_z}}{2}
    \left(\ket{3} + e^{i Q_s^b\cdot R_i }\ket{4}\right)$ is given by
\begin{align}
    E_{cl}&=E_J+E_{J'}+E_\delta \\
    \frac{1}{JN}E_J&=\frac{3}{2}(1+ P_z^2) \notag\\
    &+ \frac{1}{4}\sum_{\tau=1}^3\big( (1+ P_z)^2 \cos Q_s^t\tau + (1- P_z)^2 \cos Q_s^b\tau\notag\\
    &+ (1- P_z^2) \big[\cos Q_p\tau
    + \cos(Q_p + Q_s^b)\tau \notag \\
    &+ \cos(Q_p-Q_s^t)\tau + \cos(Q_p+Q_s^b-Q_s^t)\tau\big] \big) \\
    \frac{1}{J'N}E_{J'}&=\frac{3}{2}(1+ P_z^2) \notag\\
    &+ \frac{1}{4}\sum_{\rho=1}^3\big( (1+ P_z)^2 \cos Q_s^t\rho + (1- P_z)^2 \cos Q_s^b\rho\notag\\
    &+ (1- P_z^2) \big[\cos Q_p\rho
    + \cos(Q_p + Q_s^b)\rho \notag \\
    &+ \cos(Q_p-Q_s^t)\rho + \cos(Q_p+Q_s^b-Q_s^t)\rho\big] \big) \\
    \frac{1}{N}E_\delta&=\delta  P_z\,,
\end{align}
where $ P_z = \bra{\psi} \hat{P}_z \ket\psi$, and $\tau$ ($\rho$) are (next-)nearest-neighbor vectors of the triangular lattice. Generally, the NN sums are minimised by $Q_i=\pm K$ with $\sum_{\tau=1}^3 \cos K\tau=-3/2$, while NNN sums prefer any of the three M-vectors $Q_i=M$ since $\sum_{\rho=1}^3 \cos M \rho=-1$ while $\sum_{\rho=1}^3 \cos K\rho=3$ and $\sum_{\tau=1}^3 \cos M\tau=-1$.  

When $\delta$ is large so that $ P_z\rightarrow -1$, the leading order terms $\propto (1- P_z)^2=\mathcal O(1)$ of NN and NNN sums compete. They become equal $(E_J+E_{J'})_{Q_s^b=K}=(E_J+E_{J'})_{Q_s^b=M}$ when $J'/J=1/8$ reproducing the mean-field transition from 120$^\circ$ AFM to stripe phase in the SU(2) case. 

Thus, when we consider the case of $J'=0$ in the $ P_z\rightarrow -1$ limit, we obtain $Q_s^b=K$. Then, the next-to-leading order terms $\propto (1- P_z^2)=\mathcal O((1+ P_z))$ can all be simultaneously minimized by the configuration $Q_P=K$ and $Q_s^t=0$, which is the tripartite state we describe in the main text. The total energy of this state is $E_J/(NJ)=3(1-3 P_z)^2/8$, where $N$ is the number of lattice sites. This becomes degenerate $E_J=0$ with the manifold of three-sublattice states that have flavour-polarised sites at $ P_z=1/3$. However, to slightly increase $ P_z\gtrsim 1/3$ in the latter state, an energy of order $J$ is needed to flip one site from top to bottom layer. In contrast, the energy cost of the homogeneous configuration is much smaller $(1-3 P_z)^2\ll 1$.

Considering $J'/J>1/8$ next, the leading order term fixes $Q_s^b=M$. It is again possible to minimise all second-order terms $\mathcal O((1+ P_z))$ via $Q_s^t=M$ and $Q_P=M'$, where $M$ and $M'$ are two inequivalent M-vectors. The energy of this homogeneous four-sublattice state is $(E_J+E_J')/N=2(J+J') P_z^2$, which becomes degenerate with the four-sublattice states with flavour-polarised sites in the SU(4) limit $ P_z\rightarrow 0$. 

\section{Inclusion of quantum fluctuations}\label{appendix:inclusion}
We shall assume that, after the unitary transformation, the ground state of Eq.(\ref{eq:Heis_unit}) is very close to the fully polarised state $\prod_i \ket{1}_i$. Hence, we can use the following approximation \cite{Joshi1999,Penc2003} for the operators:
\begin{align}\label{eq:fields}
    S^1_1(i) &\sim M - \sum_{\alpha\not=1} b^\dag_\alpha(i)b^{\,}_\alpha (i), \nonumber \\
    S^{1}_\alpha(i) &\sim \sqrt{M}\,b_\alpha^\dag(i),\,\,\,\,\,\,\, (\text{with }\alpha\not= 1), \\
    S^{\alpha}_\beta(i) &\sim b_\beta^\dag(i) b^{\,}_\alpha (i) ,\,\,\,\,\,\,\, (\text{with }\alpha,\beta\not= 1), \nonumber
\end{align}
where $b_\alpha^\dag(i)$ and $b_\alpha^{\,}(i)$ are creation and annihilation operators following bosonic statistics, i.e. $[b^{\,}_\alpha(i),b^\dag_{\beta}(j)] = \delta_{ij}\delta_{\alpha\beta}$ and $M$ is the classical expectation value of $S^1_1$.
Substituting Eq.(\ref{eq:fields}) in Eq.(\ref{eq:Heis_unit}) we have:
\begin{align}\label{eq:series}
    \mathcal{U}H\mathcal{U}^\dag = H_0 + H_1 + H_2 + ...,
\end{align}

where $H_n$ contains bosonic operators to the power of $n$. Within the harmonic approximation we truncate the series in Eq.(\ref{eq:series}) to the second order.
The first term in the expansion gives back the classical energy which reads:
\begin{align}
    H_0 = NM^2 \frac{1}{2}\sum_\tau J(\tau) |\kappa_{11}(\tau)|^2 +NM\delta \tilde{P}^z_{11}.
\end{align}
It is clear from the last equation that $\delta$ must be proportional to $M$ in order to be consistent with  Eq.(\ref{eq:fields}).

The second term in the expansion contains linear terms in the bosonic fields and reads:

\begin{align}
    H_1 &=  M^{\frac{3}{2}}N\sum_{\alpha\not=1}\sum_\tau\frac{J(\tau)}{2}\left(\kappa_{11}\kappa^\dag_{1\alpha} +\kappa_{\alpha 1}\kappa^\dag_{11} \right)b_\alpha (i) + \mbox{h.c.} \nonumber \\
    &+ \delta N\sqrt{M}\sum_{\alpha\not=1}\tilde{P}^z_{\alpha 1} \,b_\alpha(i) + \mbox{h.c.}
\end{align}
This term must vanish for stability reasons and this condition fixes the value of $\delta$ that, after we set $M=1$, reads:
\begin{align}\label{eq:delta_fix}
    \delta = -\frac{\sum_\tau J(\tau) \text{Re}\, \kappa_{11}(\tau)\kappa^\dag_{1\alpha}(\tau)}{\tilde{P}^z_{1\alpha}}.
\end{align}
It is worth to note that the quantity on the right hand side of Eq.(\ref{eq:delta_fix}) does not depend on the index $\alpha$ for symmetry reasons.

After setting $M = 1$, the quadratic term reads:
\begin{align}
    H_2 &= \sum_{i}\left(-\sum_\tau J(\tau)|\kappa_{11}|^2-\delta \tilde{P}^z_{11}\right)\sum_\alpha b_\alpha^\dag(i)b^{\,}_\alpha(i)
    \nonumber \\
    &+ \sum_{i\alpha\beta}\left(\sum_\tau J(\tau) \kappa_{\alpha 1}\kappa^\dag_{\beta 1} + \delta \tilde{P}^z_{\alpha\beta}\right)b^\dag_{\beta}(i)b^{\,}_\alpha (i) \nonumber \\
    &+ \sum_{i\tau \alpha\beta}\frac{J(\tau)}{2}\,\kappa_{\alpha\beta}\,\kappa^\dag_{11}\, b^{\,}_\alpha(i)b^\dag_\beta(i+\tau) + \mbox{h.c.}
    \nonumber \\
    & + \sum_{i\tau \alpha\beta}\frac{J(\tau)}{2}\,\kappa_{\alpha 1}\,\kappa^\dag_{1\beta}\, b^{\,}_\alpha(i)b^{\,}_\beta(i+\tau) + \mbox{h.c.}
    \end{align}

where we used the short hand notation for $b_\alpha(i+\tau)$ which indicates the destruction operator of a boson with flavor $\alpha$ at site $R_i + \tau$.
After expanding the bosonic fields in their Fourier components $b_\alpha(i) =\frac{1}{\sqrt{N}}\sum_k e^{ikR_i}b_{k\alpha}$, we can rewrite the quadratic Hamiltonian in momentum space as following:
\begin{align}\label{eq:H2_ft}
    H_2 &= \sum_k\sum_{\alpha\beta}\chi_{\alpha\beta}\,b^\dag_{k\alpha}b^{\,}_{k\beta} +\left( f_{\alpha\beta}(k)b^{\,}_{k\alpha}b^\dag_{k\beta} + \mbox{h.c.}\right) + \nonumber \\
    &+\sum_k\sum_{\alpha\beta} g_{\alpha\beta}(k) b^{\,}_{k\alpha}b^{\,}_{-k\beta} + \mbox{h.c.}\, ,
\end{align}
where 
\begin{align}\label{eq:coeff:matr}
    \chi_{\alpha\beta} &= \sum_\tau J(\tau)\,\kappa_{\alpha 1}\kappa^*_{\beta 1}  \nonumber + \\&\delta \tilde{P}^z_{\alpha\beta} - \delta_{\alpha\beta}\left(\sum_\tau  J(\tau)\,|\kappa_{11}|^2 + \delta \tilde{P}^z_{11}\right) \nonumber, \\ 
    f_{\alpha\beta}(k)&= \sum_\tau \frac{J(\tau)}{2}e^{ik\tau }\kappa_{\alpha\beta}\,\kappa_{11}^*, \nonumber \\
    g_{\alpha\beta}(k) &= \sum_\tau \frac{J(\tau)}{2} e^{-ik\tau}\kappa_{\alpha 1}\,\kappa^*_{1\beta}.
\end{align}
It is useful to  introduce the set of conjugate variables $b_{k\alpha} = \frac{1}{\sqrt{2}}(x_{k\alpha}+ip_{k\alpha})$ and $b^\dag_{k\alpha} = \frac{1}{\sqrt{2}}(x_{-k\alpha}-i p_{-k\alpha})$, which obey  the canonical commuation relation $[x_{k\alpha},p_{k^\prime \alpha^\prime}] = i \, \delta_{\alpha\alpha^\prime}\delta_{k,-k^\prime}$. After substituting these expressions in Eq.(\ref{eq:H2_ft}) we  can finally write the Hamiltonian in the following matrix form:
\begin{align}\label{eq:qf}
     H_2 &= \frac{1}{2}\sum_k 
    \left(
    \begin{array}{c}
    \mathbf{p}_k\\
    \mathbf{x}_k
    \end{array}
    \right)^T
    \left[
     \begin{array}{cc}
         \mathcal{H}^P(k) &\mathcal{H}^{PX}(k)  \\
       \mathcal{H}^{XP}(k)  &  \mathcal{H}^X(k)
    \end{array}
    \right]
    \left(
    \begin{array}{c}
    \mathbf{p}_{-k}\\
    \mathbf{x}_{-k}
    \end{array}
    \right)  \nonumber \\
    &+ \mathcal{C},
\end{align}
where $\mathbf{p}_k = (p_{k1},p_{k2},p_{k3})$, $\mathbf{x}_k = (x_{k1},x_{k2},x_{k3})$, with $x_{k\alpha}$ and $p_{k\alpha}$ being conjugate variables obeying the commutation relation $[x_{k\alpha},p_{k^\prime\beta}] = i \,\delta_{k,-k^\prime}\delta_{\alpha\beta}$.
The Hamiltonian is a $6\times 6$ matrix that has been represented in a block form in Eq.(\ref{eq:qf}), where each block represents a $3\times 3$ matrix.
In particular, we have that:  
\begin{align}
        \mathcal{H}_{\alpha\beta}^P(k) &= \chi_{\alpha\beta} +f_{\alpha\beta}(k) + f_{\alpha\beta}(-k) + g_{\alpha\beta}(k) + g_{\alpha\beta}(-k) \nonumber \\
     \mathcal{H}_{\alpha\beta}^X(k) &=   \chi_{\alpha\beta} +f_{\alpha\beta}(k) + f_{\alpha\beta}(-k) - g_{\alpha\beta}(k) - g_{\alpha\beta}(-k) \nonumber \\ 
     \mathcal{H}_{\alpha\beta}^{XP}(k) &=i[-f_{\alpha\beta}(k) + f_{\alpha\beta}(-k) + g_{\alpha\beta}(k) - g_{\alpha\beta}(-k) ] \nonumber \\
      \mathcal{H}_{\alpha\beta}^{PX}(k) &=  [\mathcal{H}_{\beta\alpha}^{XP}(k)]^*,
\end{align}

The additive constant in Eq.(\ref{eq:qf}) is given by $\mathcal{C} = -\frac{N}{2}\sum_{\alpha}\chi_{\alpha\alpha}$.
The spectrum of the quantum excitations is given by the symplectic spectrum of the Hamiltonian in Eq.(\ref{eq:qf}), that coincides with the  eigenvalues $\epsilon_{k\alpha}$ of $i \mathcal{J}\mathcal{H}(k)$, with $\mathcal{J} = -i \,\sigma^{y}\otimes \mathbb{I}_{3\times 3}$ \cite{arvind1995real}.

Within the harmonic approximation, the expression of the density matrix is given by: 
\begin{align}
    n_{\alpha\beta} &= \left< \left(
    \begin{array}{c}
    \mathbf{p}_k\\
    \mathbf{x}_k
    \end{array}
    \right)^T
    \left[
    \begin{array}{cc}
    \,\,\,\ket{\alpha}\bra{\beta}& i\ket{\alpha}\bra{\beta}\\
   -i \ket{\alpha}\bra{\beta}& \ket{\alpha}\bra{\beta}
    \end{array}
    \right]
    \left(
    \begin{array}{c}
    \mathbf{p}_{-k}\\
    \mathbf{x}_{-k}
    \end{array}
    \right)
    \right>_{BZ},
\end{align}
where $\alpha,\beta = 2,3,4$ and the subscript $BZ$ means that the quantum expectation values for the different crystalline momenta must be averaged over the Brillouin zone. The expectation value of the majority flavors, according to Eq.(\ref{eq:fields}), is then given by $n_1 = 1 - \sum_{\alpha=2}^4 n_{\alpha\alpha}$.
\begin{figure}[hbtp!]
\centering
    \includegraphics[width = \columnwidth]{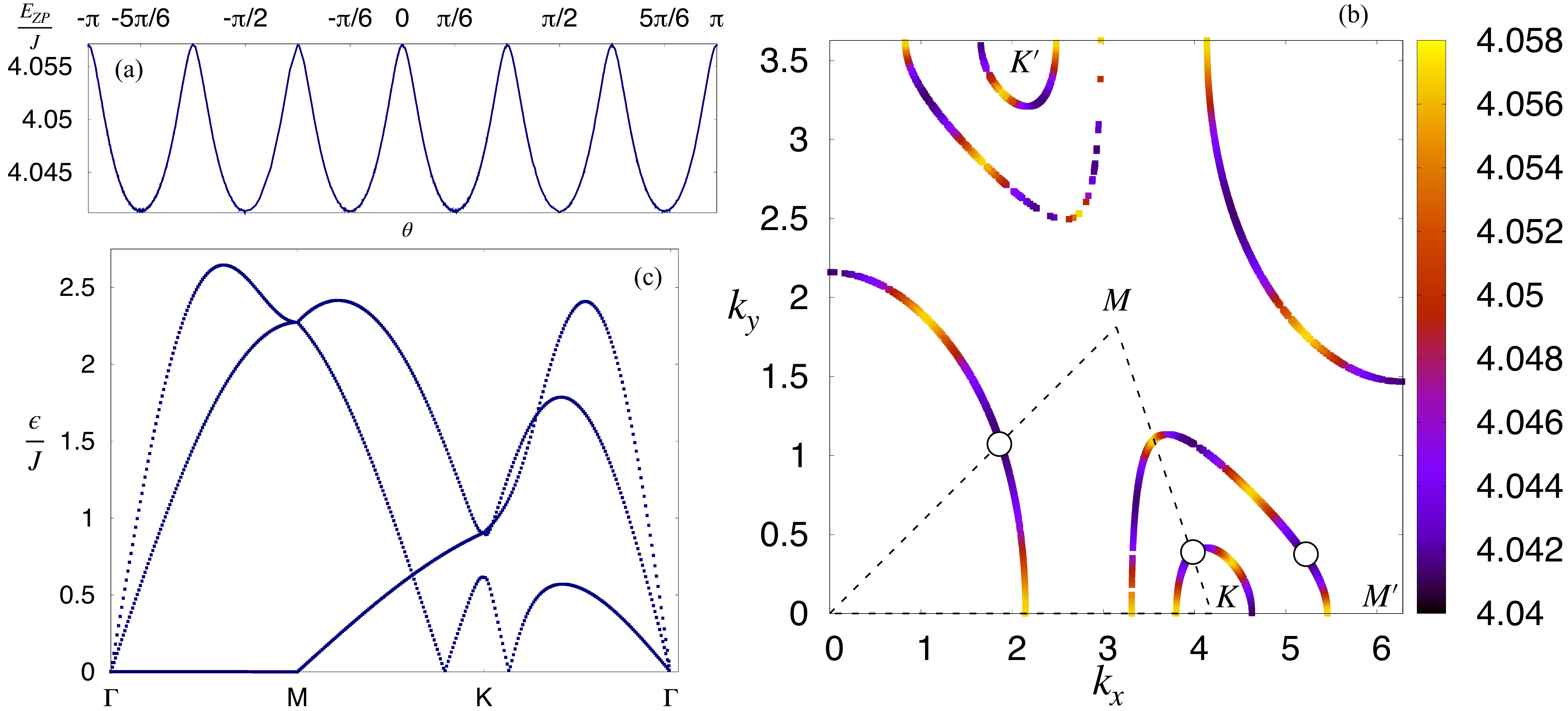}
    \caption{Quantum fluctuations for $P_z = -0.183$ which lies in the classically forbidden region. (a) Zero-point energy plotted as a function of the polar angle $\theta$. (b) Energy distribution on the classical contours shown in Figure \ref{fig:phase_diagram}. (c) Spin waves spectrum plotted as a function of the crystalline momentum.}
   \label{fig:qf_c_forb}
\end{figure}
\subsection{Order by disorder}\label{appendix:order_by_disorder}
Quantum fluctuations can remove 
the high degeneracy {of the classical ground state in the regime $|P_z|<1/3$ and} select particular states. 
{To investigate this} quantum order-by-disorder phenomenon \cite{villain1980order,Zhit2012,toth2010,Bauer2012,Jack2015} 
{we calculate} quantum corrections to the energy 
for every degenerate classical state.
This is done by evaluating the zero-point energy of the quantum fluctuations that in our case reads $E_{ZP}(\theta,\ell) = \sum_{k\alpha}\epsilon_{k\alpha}(\theta,\ell)$, where $\theta$ and $\ell = K, K^\prime$ specify the incommensurate order, and then finding its minimum as a function of the angle. In Fig.~\ref{fig:qf_c_forb} (a), we show the zero-point energy as a function of the polar angle $\theta$ for $P_z = -0.183$: the energy has six minima at $\pm \pi/6$, $\pm 5\pi /6$ and $\pm \pi /2$. 
{Since} $E_{ZP}(\theta,\ell)$ does not depend on the second index for symmetry reasons,  
every minimum is doubly degenerate.
In Fig.~\ref{fig:qf_c_forb} (b), a color plot shows how the zero-point energy is distributed on the same contours {as the ones} shown in Fig.~\ref{fig:phase_diagram}.  
The Q-vector triplet minimising the energy, given by $\theta = \pi/6$ and $\ell = K$, is plotted with white filled dots. Interestingly, $Q_s^t$ and $Q_s^b$ selected by quantum fluctuations lie respectively on the $\Gamma$-$M$ and $K$-$M$ directions. {For convenience, we show again}  
the flavor-wave spectrum as a function of the crystalline momentum for $P_z = -0.183$ in Fig.~\ref{fig:qf_c_forb}. 

\subsection{Dynamical Structure Factor}\label{appendix:dynamical}
Let us define the following retarded Green's function:
\begin{align}\label{eq:gf}
    G_{ab}(k,t) &= -i\,\theta(t)\bra{0}\left[\xi_{ka}(t),\xi_{-kb}\right]\ket{0}
\end{align}
where we introduced the six-dimensional vector $\xi_k = (\mathbf{p}_k,\mathbf{x}_k)$. We can  express 
{$G_{ab}(k,t)$} in terms of the Green's function of the quasi-particles by means of a canonical transformation that diagonalises the Hamiltonian in Eq.(\ref{eq:qf}) and that preserves the canonical variables commutation relations. {This is done via} 
a transformation $S_k$ so that $S^\dag_k \mathcal{J} S_k = \mathcal{J}$ and $S^\dag_k \mathcal{H}_k S_k = \text{diag}(\boldsymbol{\epsilon}_k,\boldsymbol{\epsilon}_k)$, where $\mathcal{H}_k$ is the 6$\times$6 matrix appearing in Eq.(\ref{eq:qf}), and $\boldsymbol{\epsilon}_k = (\epsilon_{k1},\epsilon_{k2},\epsilon_{k3})$. 
Such a transformation defines a set of new canonical coordinates, namely $\xi_k=\tilde{\xi}_{k}\, S^\dag_k$ and $\xi_{-k}=S_k\,\tilde{\xi}_{-k}$. 
In particular, let us define the following Hermitian matrix $\tilde{\mathcal{J}}_k = i\,\mathcal{H}_k^{-1/2}\mathcal{J}\mathcal{H}_k^{-1/2}$, that can be rotated into the following diagonal form $U_k^\dag\tilde{\mathcal{J}}_kU_k = \text{diag}(\boldsymbol{\epsilon}_k,-\boldsymbol{\epsilon}_k)$. It is useful to introduce the following unitary transformation $R_k = U_kT$, where $T = \exp(i \sigma^x \pi/4)\otimes \mathbbm{I}_{3\times 3}$. Finally, we can write the canonical transformation as following: $S_k = \mathcal{H}^{-1/2}_k R_k\,D_k$, with $D_k = \text{diag}(\boldsymbol{\epsilon}^{1/2}_{k},\boldsymbol{\epsilon}^{1/2}_{k})$.
\par Hence, Eq.(\ref{eq:gf}) can be rewritten in the following way:
\begin{align}\label{eq:gf1}
    G_{ab}(k,t) &= \sum_{a^\prime b^\prime} [S^\dag_k]_{a^\prime a}[S_k]_{b b^\prime} \tilde{G}_{a^\prime b^\prime}(k,t).
\end{align}
The Fourier transform of $\tilde{G}$, i.e. $\int_0^\infty \tilde{G}_{ab}(k,t)e^{i\omega t}dt $ is a 6$\times$6 matrix with the following structure:
 \begin{align}
     \tilde{G}(k,\omega)&= 
     \left(
     \begin{array}{cc}
 \delta_{\alpha\beta}G_{\alpha}^X(k,\omega)& -\delta_{\alpha\beta}G_{\alpha}^{XP}(k,\omega)\\
  \delta_{\alpha\beta}G_{\alpha}^{XP}(k,\omega)& \delta_{\alpha\beta}G_{\alpha}^X(k,\omega)
     \end{array}
     \right),
 \end{align}
 where $\alpha = \{1,2,3\}$ and  the block diagonal terms are given by:
 \begin{align}\label{eq:gf2}
     G_{\alpha}^X(k,\omega) &= \frac{1}{2}\left(\frac{1}{\omega - \epsilon_{k\alpha} + i \eta} - \frac{1}{\omega + \epsilon_{k\alpha}  + i \eta}\right), \nonumber \\
     G_{\alpha}^{XP}(k,\omega) &=  \frac{i}{2}\left(\frac{1}{\omega -\epsilon_{k\alpha} + i\eta}+  \frac{1}{\omega + \epsilon_{k\alpha}  + i \eta}\right).
 \end{align}
 The Green's function evaluated along the imaginary Matsubara frequencies can be calculated starting from the one  in Eq.(\ref{eq:gf2}) via analytic continuation, i.e. by replacing $\omega + i\eta \to i\omega_n$, with $\omega_n = \frac{2\pi n}{\beta}$.\\

 Since, flavor-flip processes where the majority flavor $\ket{1}$ flips into  the minority ones ($\ket{\alpha} = $ $\ket{2}$,$\ket{3}$,$\ket{4}$) are encoded by the operators  $S^1_\alpha $, the dynamic structure factor can be written as:
\begin{align}
   \mathcal{C}(k,\omega) &= -\,\text{Im}\,i\int_0^\infty dt \,e^{-i\omega t}\sum_\alpha
     \bra{0}[S^\alpha_{1}(k,t), S^1_\alpha(-k)]\ket{0}
   \nonumber \\
   &=-\,\text{Im}\,i\int_0^\infty dt \,e^{-i\omega t}\sum_\alpha
     \bra{0}\big([x_{k\alpha}(t), x_{-k\alpha}] \nonumber \\
   &+ [p_{k\alpha}(t), p_{-k\alpha}] + i [p_{k\alpha}(t), x_{-k\alpha}] 
   \nonumber \\ 
   &- i [x_{k\alpha}(t), p_{-k\alpha}]
   \big)\ket{0} \nonumber \\
   &= -\text{Im}\,\text{Tr} \, \left[(\mathbbm{I}_{6\times 6} + \sigma^y\otimes\mathbbm{I}_{3\times 3})G(k,\omega)\right].
    \end{align}

\begin{figure}
    \centering
    \includegraphics[width = \columnwidth]{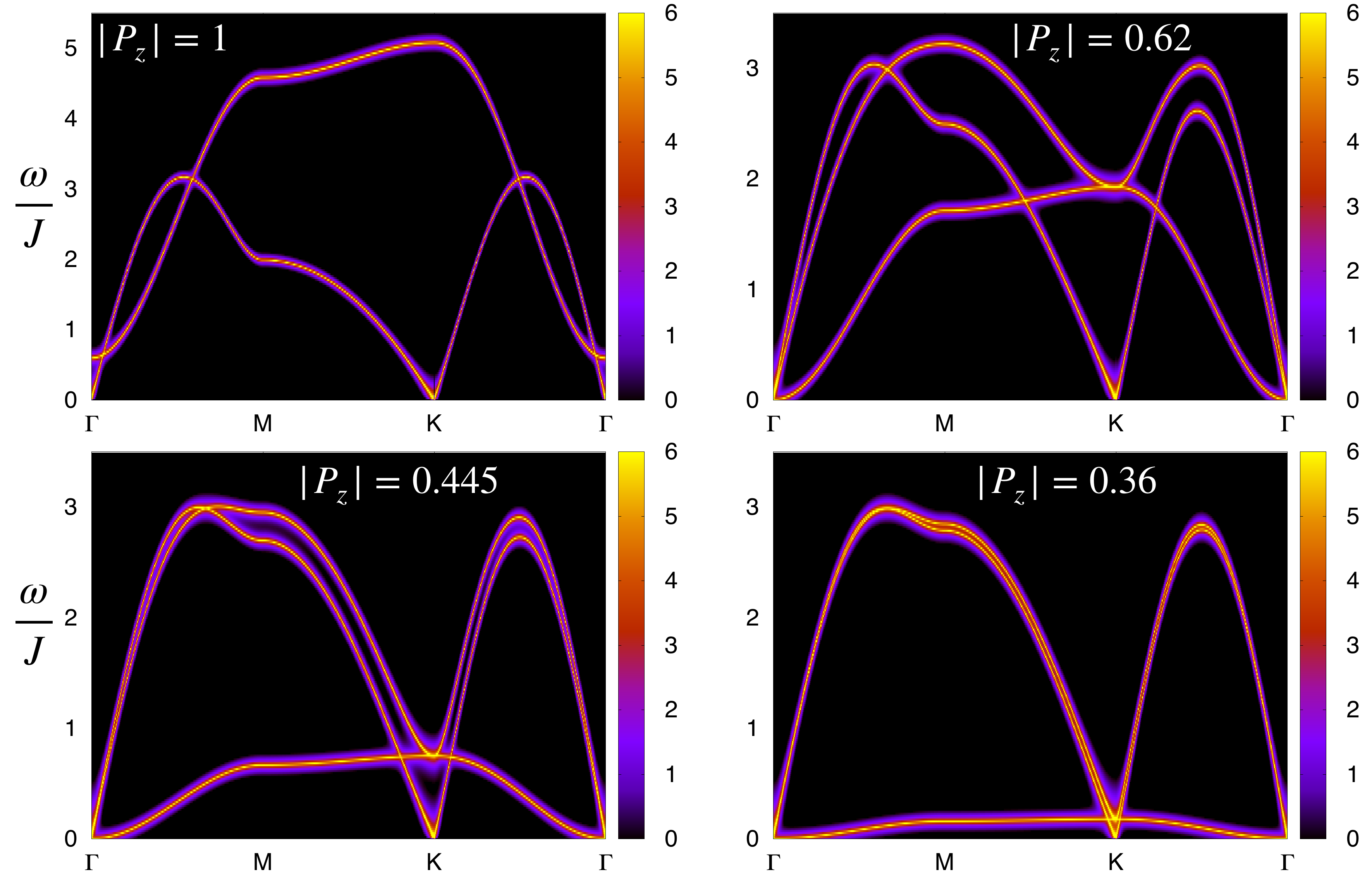}
    \caption{Dynamic structure factor plotted in log scale for different polarisation values $|P_z| = 1, 0.62, 0.445, 0.36$ for $J^\prime = 0$. The ferromagnetic spin-wave mode that flattens to zero at $|P_z|=1/3$ is well visible for all values of the layer imbalance.}
    \label{fig:spectral}
\end{figure}

Let us note that, since the canonical transformation $S_k$ is not unitary, the trace appearing in the last equation is affected non trivially by the weights $[S_k]_{ab}$ in Eq.(\ref{eq:gf1}). For this reason it is 
{useful} to evaluate the dynamical structure factor which measures the absorption intensity.
 Fig.~\ref{fig:spectral} shows the dynamic structure factor for different values of the layer polarisation and for $J^\prime = 0$. {We clearly recover the flavour-wave spectrum. In particular,} we observe that the ferromagnetic mode that flattens to zero at $|P_z| = 1/3$ is  clearly visible for all values of the polarisation. Therefore, the softening of this mode could be used as direct experimental evidence of the high-degeneracy of the excitation spectrum which we identified as the main mechanism leading to the suppression of LRO and the onset of possible spin-liquid phases. 
\clearpage
\bibliography{biblio.bib}
\end{document}